\documentclass{PoS}
\title{A new method of calculating the running coupling constant 
\footnote{Based on the contributions of E. Itou and M. Kurachi}}

\ShortTitle{A new method of calculating the running coupling constant }

\author{Erek~Bilgici$^a$, Antonino~Flachi$^b$, Etsuko~Itou$^{b,\dag}$, Masafumi~Kurachi$^{b,\dag}$, C.-J~David~Lin$^{c, d}$, Hideo~Matsufuru$^e$, Hiroshi~Ohki$^{b, f}$, Tetsuya~Onogi$^b$ and~Takeshi~Yamazaki$^b$
\\
$(a)$ Institut f\"{u}r Physik, Universit\"{a}t Graz, A-8010 Graz, Austria\\
$(b)$ Yukawa Institute for Theoretical Physics, Kyoto University, Kyoto 606-8502, Japan\\
$(c)$ Institute of Physics, National Chiao-Tung University, Hsinchu 300, Taiwan\\
$(d)$ Physics Division, National Centre for Theoretical Sciences, Hsinchu 300, Taiwan\\
$(e)$ High Energy Accelerator Research Organization (KEK), Tsukuba 305-0801, Japan\\
$(f)$ Department of Physics, Kyoto University, Kyoto 606-8501, Japan\\
$^\dag$ speaker\\
E-mail: \email{itou@yukawa.kyoto-u.ac.jp},
\email{kurachi@yukawa.kyoto-u.ac.jp}
}

\abstract{
We propose a new method to compute the running coupling constant
of gauge theories on the lattice. We first give the definition of the
running coupling in the new scheme using the Wilson loops in a finite
volume, and explain how the running of the coupling constant is
extracted from the measurement of the volume dependence. The perturbative
calculation of the renormalization constant to define the scheme is also
given at the leading order.  As a benchmark test of the new scheme 
we apply the method the case of the quenched QCD. We show
the preliminary result from our numerical simulations which are carried out
with plaquette gauge action for various lattice sizes and bare lattice couplings.
With techniques to improve the statistical accuracy, we show that we can
determine the non-perturbative running of the coupling constant in a wide
range of the energy scale with relatively small number of gauge configurations
in our scheme. We compare our lattice data of the running coupling constant
with perturbative renormalization group evolution at one- and  two-loop order,
and confirm the consistency between them at high energy.
}

\FullConference{The XXVI International Symposium on Lattice Field Theory \\
		 July 14 - 19, 2008\\
		 Williamsburg, Virginia, USA\\
Report number: YITP-08-68}

\newcommand{\beq}{\begin{eqnarray}}
\newcommand{\eeq}{\end{eqnarray}}
\begin{document}

%%%%%%%%%%%%%%%%%%%%%%%%%%%%%%%%%%
\section{Introduction}

The properties of a vectorial gauge theory which has (near) conformal 
infrared fixed point (IRFP) are of fundamental importance. In addition to its 
intrinsic field-theoretic interest, this approximate conformal, or ``walking'', 
behavior is an essential ingredient of modern technicolor models of 
dynamical electroweak symmetry breaking \cite{tc}, providing requisite 
enhancement of Standard Model fermion masses \cite{wtc1, wtc2, 
chipt1, chipt2, my, chipt3}. This walking  behavior could also give a 
remedy for the problem of large correction (often denoted as $S$ 
\cite{pt, ab}) to the $Z$ boson propagator  \cite{Appelquist:1998xf, 
Harada:2005ru, Kurachi:2006mu}. Our ultimate objective is to investigate 
the nature of physics in such a near conformal gauge theory.
However, we first have to address the following question: is there any theory 
which actually has a conformal nature in a certain energy region? 

An $SU(N)$ gauge theory with a large number of massless fermions 
has been known as a promising candidate for a theory which has the 
nontrivial IRFP \cite{bz}. For example, in the case of $SU(3)$ 
gauge theory, 
%(in which case, it is often called the large $N_f$ QCD), 
when the number of massless fermions, $N_f$, takes values in 
the range $8 < N_f \le 16$, the two-loop running coupling approaches 
to a finite value, 
$\alpha_\ast = \frac{4\pi (33/2 - N_f)}{19(N_f - 153/19)}$, 
in the infrared energy region due to the existence of the IRFP.
When $N_f$ is close to $16$ (which is the largest $N_f$ we can take 
when we restrict our study to asymptotically free theories), 
even though the value of $\alpha_\ast$ is corrected by  
higher order terms in the perturbative expansion, the existence of 
the IRFP is valid beyond two-loop approximation since the coupling is 
small enough. However, when 
we consider a range of $N_f$ where the above two-loop estimation of 
$\alpha_\ast$ becomes large, it is quite non-trivial whether this IRFP 
really exists beyond perturbation theory. 

In principle, lattice simulations should provide a way to determine 
whether an $SU(N)$ gauge theory with a certain number of 
massless fermions has conformal 
fixed point or not. Several pioneering works \cite{lgt} found the chiral phase 
transition at certain critical flavor, which can be considered as an indication 
of the appearance of the IRFP. However, groups that have studied this have not 
reached consensus on the critical number of flavor for the chiral phase 
transition. Recently, Appelquist et al.~\cite{Appelquist:2007hu} 
directly calculated the running coupling of the 
$SU(3)$ gauge theory with $8$ and $12$ flavors by using the Schr\"{o}dinger 
functional (SF) method on the lattice, and showed 
an evidence for the existence of the IRFP for $N_f=12$ while the 
running coupling for $N_f=8$ is more like QCD with three flavors.

Here, we propose a new scheme for the calculation of the running 
coupling on the lattice. 
We extract the renormalized coupling from the Wilson loops in a finite
volume, and determine the running of the coupling constant 
from the measurement of the volume dependence 
by using the step scaling procedure. This new scheme can serve as 
an alternative computational method in general gauge theories.

In the context of exploring the conformal window, 
using a different scheme to calculate the coupling constant
is well-motivated by itself to verify the real existence of the fixed point.  
Technical advantage of our scheme is that it is free from $O(a)$ 
error in contrast to the SF scheme, in which it is practically difficult
to remove all the $O(a)$ discretization effects.

In Section $2$, we define the new scheme for the determination 
of the running coupling. Then we review the step-scaling procedure 
to show how to actually obtain the evolution of the running coupling 
from lattice simulations in Section $3$. Sections $4$ and $5$ are devoted 
to the explanation of the 
details of our numerical simulations, and  in Section $6$, we show the  
result of the quenched study as a test to see the effectiveness of our 
new scheme for the calculation of the running coupling. Discussion 
on the numerical results are also given in Section $6$. 
Section $7$ summarizes our conclusions.

%%%%%%%%%%%%%%%%%%%%%%%%%%%%%%%%%%
\section{Wilson Loop Scheme}

\begin{figure}[t]
   \begin{center}
     \includegraphics[height=5cm]{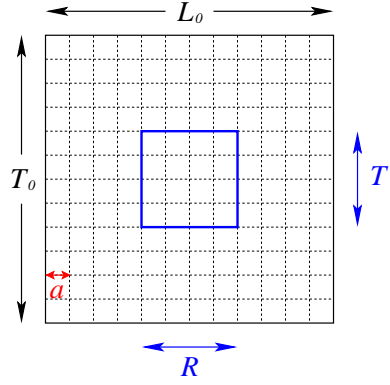}
   \end{center}
 \caption{Wilson loop defined on the latticized space-time box. $T_0$,  $L_0$ 
 and $T$, $R$ represent the size of the box and the Wilson loop in the temporal 
 and spatial directions, respectively. $a$ is the lattice spacing.}
 \label{fig:Wloop}
 \end{figure}

\begin{figure}[t]
\begin{center}
\unitlength=1mm
\begin{picture}(120,70)
     \put(78,58.2){$L_0/a = $}
     \put(0,38){\Large $k$}
     \put(55,0){\large $R/L_0$}
     \includegraphics[height=7cm]{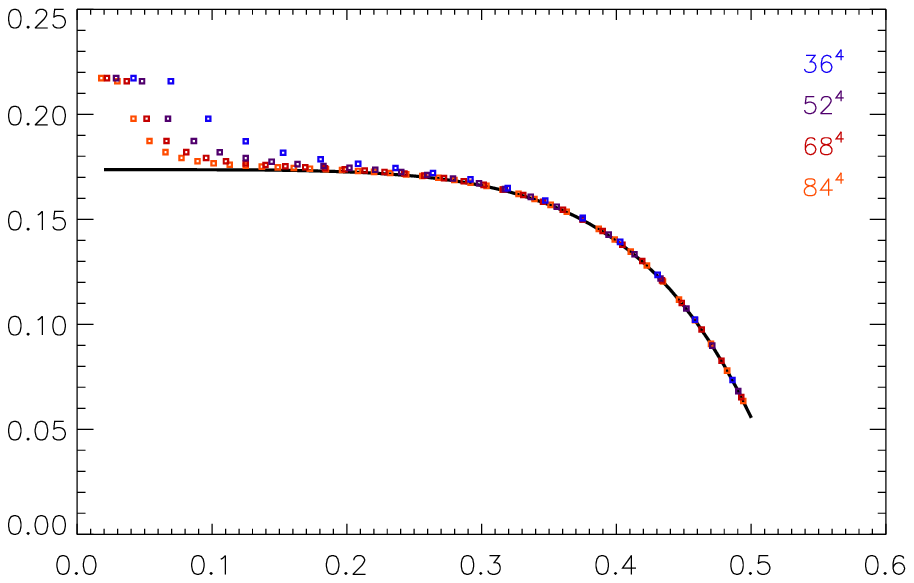}
 \end{picture}
     \caption{Values of $k$ for several values of $R/L_0$ and $L_0/a$ 
 (colored squares whose $L_0/a$ is indicated by the numbers with the same color). 
 The value of $k$ in the continuum limit is also shown as a solid curve.}
 \end{center}
 \label{fig:k}
 \end{figure}

In this section, we give the definition of the new renormalization scheme, 
the ``Wilson loop scheme'', for the running coupling, and show how to calculate it on the lattice.
Let us start with general features in the renormalization of the coupling constant. 
 Consider an amplitude $A$ whose tree-level contribution 
is proportional to $g_0^2$ (where $g_0$ is the bare coupling constant):
\begin{equation}
 A^{\rm tree} = k g_0^2 .
 \label{eq:tree}
\end{equation}
Here, $k$ is a certain coefficient which is a function of all the parameters 
of the theory except $g_0$. Then, we denote the ratio of the fully 
non-perturbative value of the amplitude $A$ to its tree-level value as 
$Z(\mu)$:
\begin{equation}
 A^{\rm NP}(\mu) = Z(\mu) A^{\rm tree}, 
\end{equation}
where $\mu$ is the scale at which the amplitude $A$ is defined. 
By using Eq.~(\ref{eq:tree}), the right hand side of the above equation 
can be rewritten as $Z(\mu) g_0^2\ k$, and the combination 
$Z(\mu) g_0^2$ can be identified as the renormalized coupling 
at the scale $\mu$. So the renormalized coupling, $g(\mu)$, can be 
expressed as follows:
\begin{equation}
 g^2(\mu) = \frac{A^{\rm NP}(\mu)}{k} .
 \label{eq:g-mu}
\end{equation}
Thus, any amplitude with a tree-level value proportional to $g_0^2$ 
can be used to define the renormalized coupling. 
Here, we use the following quantity:
\begin{equation}
 - R^2 \left. \frac{\partial^2}{\partial R \partial T}\ln \langle W(R, T; L_0, T_0) \rangle \right|_{T=R}, 
\end{equation}
where $W(R, T)$ is the Wilson loop. The definition of the Wilson loop 
is graphically shown in Fig.~\ref{fig:Wloop}. 
In this figure, 
$T_0$,  $L_0$, and $T$, $R$ represent the size of the box and the Wilson loop 
in the temporal and spatial directions, respectively, and $a$ is the lattice spacing.
From now on, for simplicity, we consider the case of $T_0 = L_0$.
At tree level in the perturbative expansion, this quantity actually is proportional 
to $g_0^2$, \textit{i.e.}, 
\begin{equation}
 - R^2 \left. \frac{\partial^2}{\partial R \partial T}\ln \langle W(R, T; L_0) \rangle^{\rm tree} \right|_{T=R} 
 = k g_0^2,  
\end{equation}
where $k$, in the case of periodic boundary condition for example, 
is  
\begin{eqnarray}
k &=& 
 -  R^2  \frac{\partial^2}{\partial R \partial T} \left[\ \ 
 \frac{4}{(2\pi)^4}
\sum_{n_0, n_1, n_2, n_3 (\neq 0)} 
\left(
 \frac{\sin(\frac{\pi n_0 T}{L_0})}{n_0}
\right)^2
\frac{e^{i\frac{2 \pi n_1 R}{L_0}}}{n_0^2 + \vec{n}^2}\ \ \ 
\right]_{T=R}  \nonumber \\
& & + \ {\rm zero\ mode\ contribution}. 
\label{eq:k}
\end{eqnarray}
Here, $(n_0, n_1, n_2, n_3)$ represents integer four-vector to define 
the momentum. 
``Zero mode contribution'' in the above equation is coming from 
the existence of so-called ``toron'' contributions which originate from 
zero-mode configurations degenerate with the vacuum on the periodic torus. 
This contribution is calculated 
in Ref.~\cite{Coste:1985mn}, and we use the result from that 
paper. After evaluating the summation in Eq.~(\ref{eq:k}) \footnote{
Detailed calculation of the factor $k$ can be found in \cite{k-paper}.}, 
one can find that $k$ only depends on the value of $R/L_0$.
The value of $k$ as a function of $R/L_0$ in the continuum limit is shown in Fig.~\ref{fig:k}. 
We also did similar calculations of $k$ in the case of discrete space-time, 
and plotted them for several values of $L_0/a$ and $R/L_0$. Note that the 
continuum limit actually exists (\textit{i.e.}, $k$ is finite in the limit of $L_0/a \rightarrow \infty$) 
and that the convergence to continuum value is faster for larger values of $R/L_0$. 
Once the value of $k$ is obtained, the renormalized coupling is defined according to  
Eq.~(\ref{eq:g-mu}):
 \begin{equation}
   g^2\left(L_0, \frac{R}{L_0}\right) \ \ = \ \ 
   \frac{ \ - R^2 \left. \frac{\partial^2}{\partial R \partial T}\ln 
   \langle W(R, T; L_0) \rangle^{\rm NP} \right|_{T=R}    \ }
   {k\left(\frac{R}{L_0}\right)}.
   \label{eq:g-L0}
 \end{equation}
The numerator on the right hand side of 
Eq.~(\ref{eq:g-L0}) can be estimated from the Creutz ratio on the lattice, 
\begin{equation}
 \chi(\hat{R}+1/2, \hat{T}+1/2;  L_0/a)  = 
 - \ln 
 \left(  
 \frac{W(\hat{R}+1, \hat{T}+1;  L_0/a)\ W(\hat{R}, \hat{T};  L_0/a)}
 {W(\hat{R}+1, \hat{T};  L_0/a)\ W(\hat{R}, \hat{T}+1;  L_0/a)}
 \right),
\end{equation}
where $\hat{T} \equiv T/a$ and $\hat{R} \equiv R/a$.
The value of $\chi$ is evaluated by a Monte Carlo (MC) simulation.
Then the renormalized coupling constant in the Wilson loop scheme can be written as:
\beq
g_{w}^2 \left(L_0, \frac{R+a/2}{L_0},\frac{a}{L_0} \right) = 
(\hat{R}+1/2)^2 \cdot \chi(\hat{R}+1/2;  L_0/a)/k, 
\label{eq:def-g-wilson}
\eeq
where we used the shorthand notation  
$\chi(\hat{R}+1/2, \hat{T}+1/2;  L_0/a)\vert_{\hat{R}=\hat{T}} \equiv \chi(\hat{R}+1/2;  L_0/a)$. 
Here, since $g_w^2$ depends on three different scales, namely, $L_0$, $R$, 
and $a$, we indicated that $g_w^2$ can be viewed as a function of $L_0$, $(R+a/2)/L_0 (\equiv r)$, 
and $a/L_0$. We choose a specific value of $r$ ($r=0.3$, for example) and 
keep it fixed to that value throughout the analysis. Varying $r$ 
means changing renormalization scheme. As for $a/L_0$, we extrapolate 
it to zero when taking the continuum limit.
After fixing these two dimensionless parameters $r$ and $a/L_0$, 
$g_w^2$ becomes a function of only one scale, $L_0$. In our scheme, $L_0$ 
is identified as the scale at which the renormalized coupling is defined. 

One thing we should emphasize here is that the Creutz ratio is free from 
$O(a)$ discretization error, mainly because $O(a)$-improvement of the heavy quark propagator is automatically achieved after the redefinition 
of the mass and the wavefunction \cite{Necco:2001xg}. 
Thus, our scheme explained here does not 
have any $O(a)$ systematic error as long as we use actions which 
do not have $O(a)$ error.

%%%%%%%%%%%%%%%%%%%%%%%%%%%%%%%
%%%%%%%%%%%%%%%%%%%%%%%%%%%%%%%

\section{Step scaling}
Here, we review the step-scaling procedure \cite{Luscher:1991wu, Caracciolo:1994ed} 
in the Wilson loop scheme, which enables us to evaluate the evolution of the running coupling 
for a large range of energy scale on the lattice.
First, we choose a specific value for $g_{w}^2$, $g_w^2=\tilde{g}_w^2$.
Then, for a fixed value of $r$, we find sets of parameters, $(\beta, L_0/a)$, 
which reproduce $\tilde{g}_w^2$ for several different values of $L_0/a$:
\beq
\left\{ \left( \beta^{(1)}_1, \left( L_0/a \right)^{(1)}_1 \right),\left( \beta^{(1)}_2, \left( L_0/a \right)^{(1)}_2 \right),\cdots \right\}.
\eeq 
What we are doing here is tuning the value of $\beta$ in such a way that the 
physical volume $L_0$ is fixed for different values of $L_0/a$. 
Let us call this fixed physical volume for the starting point as $\tilde{L}_0$, \textit{i.e.},\ \ 
$g_w^2(\tilde{L}_0) = \tilde{g}_w^2$\,. Next thing to do is to vary the physical volume 
from $\tilde{L}_0$ to $s \tilde{L}_0$, which gives the evolution of 
the running coupling from the energy scale $1/\tilde{L}_0$ to $1/s\tilde{L}_0$.
Here $s$ is the scaling factor. 
This step can be achieved by changing the lattice size from $(L_0/a)^{(1)}$ to 
$s (L_0/a)^{(1)}$ with each value of $\beta^{(1)}$ unchanged.
Values of $g_w^2$ calculated with these new parameter sets should be considered as the 
coupling at the energy scale $1/s\tilde{L}_0$ up to discretization error, so the 
extrapolation to the continuum limit can be taken by using those data. 
\beq
g^2_{R}\left( \frac{1}{s\tilde{L}_0}\right) \equiv 
\lim_{a \rightarrow 0}\left[
Z  \left(\frac{1}{s\tilde{L}_0}, \frac{a}{s\tilde{L}_0} \right)  g_{0}^2(a) 
\right] .
\eeq
The  resultant value of coupling, $g_R^2$, should be considered as the 
renormalized coupling at the energy scale $1/s\tilde{L}_0$. This is the way to 
obtain a single discrete step of evolution of the running coupling with the scaling factor $s$.
 
Next, we find new parameter sets of $(\beta^{(2)}, (L_0/a)^{(2)})$ which reproduce the value of $g_w^2(s\tilde{L}_0)$ obtained in the previous step. 
Here, we find these parameter sets in such a way that the new lattice size $L_0/a^{(2)}$ is equal to the original small lattice $(L_0/a)^{(1)}$.
From here, we can repeat exactly the same procedure as that of the first step scaling explained in the previous paragraph: we calculate $g_w^2$ with the parameter sets $(\beta^{(2)}, s(L_0/a)^{(1)})$.
By iterating this procedure, say, $n$ times, we obtain the evolution 
of the running coupling from the energy scale $1/\tilde{L}_0$ to 
$1/(s^n\tilde{L}_0)$. 

In our exploratory study, we slightly modify the step scaling 
procedure. Instead of tuning $\beta$ to keep $g_w^2(\tilde{L}_0)$ fixed, we use the 
results given by Alpha collaboration \cite{Capitani:1998mq}, namely, we use the parameter 
sets which are tuned to keep the renormalized coupling in the SF scheme, 
$g_{SF}^2(\tilde{L}_0)$, fixed. 
We also use the result given in \cite{Guagnelli:2004za} to obtain 
additional parameters for the study of lower energy region.

%%%%%%%%%%%%%%%%%%%%%%%%%%%%%%%%%%
\section{Simulation parameters}
We consider a four-dimensional Euclidean lattice $L_0^4$.
In this work, we use the standard Wilson plaquette gauge action.
Gauge configurations are generated by the pseudo-heatbath algorithm and over-relaxation, 
mixed in the ratio of $1$:$5$. We call the combination of one pseudo-heatbath update sweep 
followed by five over-relaxation sweeps ``an iteration''.
In order to ensure decorrelation, we save gauge configurations separately by $1000$ 
iterations and perform measurements on them. The number of gauge configurations we generated is 
$100$ at each set of $(\beta, L_0/a)$. 
We use both periodic and twisted boundary conditions. 
However, in this article we only report the case of periodic boundary condition.

We have five parameter sets (Set~$1$- Set~$5$) of $(\beta, L_0/a)$ corresponding to the 
fixed physical box size $L_0$ as in Table \ref{table:parameter-set}.
%%%%%%%%%%%%%%%%%%%%%%%%%%%%%%%%%%%%%%%%%%%%%%%%%%%%%%%
%%%%%%%%%%%%%%%%%%%%%%%%%%  TABLE  %%%%%%%%%%%%%%%%%%%%
%%%%%%%%%%%%%%%%%%%%%%%%%%%%%%%%%%%%%%%%%%%%%%%%%%%%%%%
	
	\begin{table}[h]
	
	\begin{center}
	\begin{tabular}{|c|c|c||c|c|c||c|c|c||c|c|c|}
	\hline
Set~$1$ & &  & Set~$2$ &&  & Set~$3$ &&  & Set~$4$ && \\
\hline \hline
$\beta$ & $L_0/a$& $L_0/a$& $\beta$& $L_0/a$& $L_0/a$& $\beta$ & $L_0/a$ & $L_0/a$& $\beta$ & $L_0/a$ & $L_0/a$  \\
{} & \footnotesize{$(s=1)$}& \footnotesize{$(s=2)$}&
{} & \footnotesize{$(s=1)$}& \footnotesize{$(s=2)$}&
{} & \footnotesize{$(s=1)$}& \footnotesize{$(s=2)$}&
{} & \footnotesize{$(s=1)$}& \footnotesize{$(s=2)$}\\
\hline
  $8.2500$   & $(8)$  & $16$ & $7.6547$   & $(8)$  & $16$
& $7.0197$   & $(8)$  & $16$ & $6.4527$   & $(8)$  & $16$\\
  $8.4677$   & $(10)$ & $20$ & $7.8500$   & $(10)$ & $20$ 
& $7.2098$   & $(10)$ & $20$ & $6.6629$   & $(10)$ & $20$ \\
  $8.5985$   & $12$   & $24$ & $7.9993$   & $12$   & $24$
& $7.3551$   & $12$   & $24$ & $6.7750$   & $12$   & $24$\\
  $8.7289$   & $14$   &    & $8.1352$   & $14$   &    
& $7.4986$   & $14$   &    & $6.9169$   & $14$   &     \\
  $8.8323$   & $16$   &    &$8.2415$   & $16$   &   
& $7.0203$   & $16$   &    &$7.6101$   & $16$   &       \\
	\hline
	\end{tabular}
\vspace{1cm}

\begin{tabular}{|c|c|c|c|}
\hline
Set~$5$ & {} & {} & {} \\
\hline	\hline
$\beta$      & $L_0/a$ & $L_0/a$ & $L_0/a$ \\
{} & \footnotesize{$(s=1)$}& \footnotesize{$(s=1.5)$} &\footnotesize{$(s=2)$}\\
\hline
$6.1274$  & (8)     &  12     &   16    \\
$6.2647$  & (10)    &         &   20    \\
$6.3831$  & 12      &  18     &   24    \\
$6.4841$  & 14      &         &         \\
$6.5700$  & 16      &  24     &         \\
	\hline
\end{tabular}
	\caption{The parameter sets of $\beta$ and $L_0/a$ used for measurements. 
	Each of the first columns in Sets~$1$-$4$ 
	gives the constant SF coupling, and the first one in Set~$5$ gives the constant Sommer scale. 
	The parameter sets of $L_0/a=8$ and $10$ (denoted by parentheses) are used only for the 
	reference to set the scale, but not for measurements.
	}
	\label{table:parameter-set}
	\end{center}
	\end{table}
	
%%%%%%%%%%%%%%%%%%%%%%%%%%%%%%%%%%%%%%%%%%%%%%%%%%%%%%%%
%%%%%%%%%%%%%%%%%%%%%%%%%%%%%%%%%%%%%%%%%%%%%%%%%%%%%%%%
%%%%%%%%%%%%%%%%%%%%%%%%%%%%%%%%%%%%%%%%%%%%%%%%%%%%%%%%
The values of $\beta$ for Set~$1$-$4$ are tuned by fixing $L_0$ using the renormalized 
coupling in SF scheme, $g_{SF}^2$, as inputs. (See, Table~$6$ in \cite{Capitani:1998mq}.) 
Also, these parameters are tuned in such a way that the value of $g_{SF}^2$ obtained 
from the first column (\textit{i.e.} $s=1$) of Set~$2$ (or $3$, $4$) coincide with that obtained from 
the second column (\textit{i.e.} $s=2$) of Set~$1$ (or $2$, $3$) up to systematic and statistical errors. 
Thus, if we call the energy scale defined by the $s=1$ of Set~$1$ as $1/\tilde{L}_0$, from the 
simulations with parameters in Set~$1$-$4$, we obtain the evolution of the running coupling 
from $1/\tilde{L}_0$ to $1/(2^4 \, \tilde{L}_0)$. We also added an extra set (Set~$5$) of 
parameters to obtain further step scalings toward low energy region. Parameters in 
Set~$5$ are tuned by using the Sommer scale, $r_0$, as inputs. 
(See Eq.~($2.18$) in \cite{Guagnelli:2004za}.)

Note that even though we use existing parameter sets which are tuned by using $g_{SF}^2$ or 
$r_0$ as an input here, to make our study self-contained, we should 
tune parameters by using $g_w^2$ itself. Such study is underway, and will be presented 
in our future publication~\cite{k-paper}.

%%%%%%%%%%%%%%%%%%%%%%%%%%%%%%%%%%%%%%%%%%%%%%%%%%%%%%%%
%%%%%%%%%%%%%%%%%%%%%%%%%%%%%%%%%%%%%%%%%%%%%%%%%%%%%%%%
%%%%%%%%%%%%%%%%%%%%%%%%%%%%%%%%%%%%%%%%%%%%%%%%%%%%%%%%

%%%%%%%%%%%%%%%%%%%%%%%%%%%%%%%%%%
\section{Simulation details}
Now, we calculate the renormalized coupling constant in the 
Wilson loop scheme defined by Eq.~(\ref{eq:def-g-wilson}) for a fixed value of $r$:
\beq
g_{w}^2 (L_0, a/L_0) =(\hat{R}^2+1/2)^2\cdot \chi(\hat{R}+1/2; L_0/a) /k|_{\mbox{\tiny{fixed $r$}}}.
\eeq

\begin{figure}[h]
   \begin{center}
   \unitlength=1mm
   \begin{picture}(105,75)
     \put(-2,72){\large $(\hat{R}+1/2)^2 \cdot \chi$}
     \put(49,-2){\large $\hat{R}+1/2$}
     \includegraphics[height=7cm]{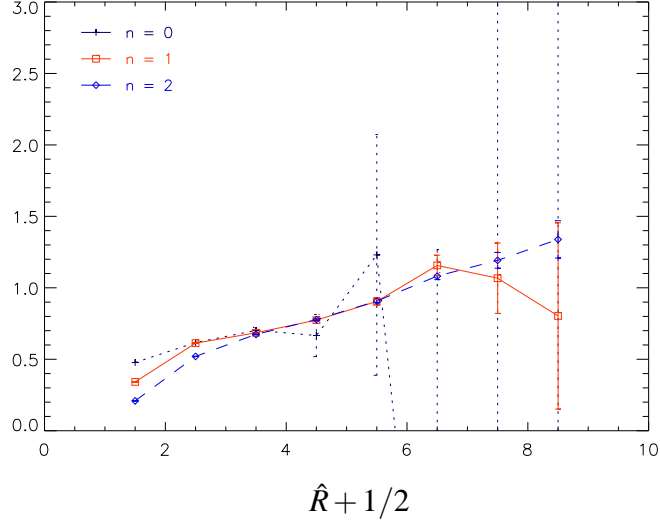}
     \end{picture}
   \end{center}
 \caption{The values of $(\hat{R}+1/2)^2 \cdot \chi$ with statistical error for  several $(\hat{R}+1/2)$ 
 in the case of $\beta = 6.3831$ and $L_0/a = 18$. The black cross, the red square and the blue diamond denote the data with $0$, $1$ and $2$ smearing, respectively.}
 \label{fig:Smear-Plot}
 \end{figure}

First, to reduce the statistical error, we use the APE smearing \cite{Albanese:1987ds} of link variables defined by the following equation;
\beq
U^{(n+1)}_{x,\mu}=Proj_{SU(3)} \left[ U^{(n)}_{x,\mu}+\frac{1}{c} \Sigma^{4}_{\mu \ne \nu}U^{(n)}_{x,\nu} U^{(n)}_{x+\nu,\mu} U^{(n) \dag}_{x+\mu,\nu} \right],
\eeq
where $n$ and $c$ denote a smearing level and a smearing parameter, respectively.
The result does not depend on the value of $c$ significantly, and we take $c=2.3$ in this work.
Note that we need to find a optimal set of $r \equiv \frac{R+a/2}{L_0}$ and the smearing level $n$ 
by considering the following requirements. 
To control the discretization error, it is better to choose a larger $r$.
For the purpose of reducing the statistical error, it is better to take a smaller $r$ and higher $n$. 
Fig.~\ref{fig:Smear-Plot} shows the  smearing-level dependence of 
$(\hat{R}+1/2)^2 \cdot \chi$ in the case of $\beta=6.3831$ and $L_0/a=18$ as an example.
From this figure, we find the statistical error is notably reduced even at smearing level one.
Furthermore, in order to avoid over-smearing, $n$ should be smaller than $\hat{R}/2$. 
This condition gives the lower bound, $L_0/a > (4n+1)/2$. 
We summarize the bound from this requirement in Table~\ref{table:smearing-level}.
We actually see, for example in the case of $L_0/a=18$ (see Fig.~\ref{fig:Smear-Plot}), 
that the data of $(\hat{R}+1/2)=1.5$, $2.5$ in higher smearing level are not reliable because of 
over-smearing. By considering all of the above requirements, we find that $(r,n)=(0.3, 1)$ is 
an optimal choice.

%%%%%%%%%%%%%%%%%%%%%%%
%%%%%%%%%%%%%%%%%%%%%%%

%%%%%%%%%%%%%%%%%% Table %%%%%%%%%%%%%%%%%%%%%%
\begin{table}[h]
	
	\begin{center}
	\begin{tabular}{|c|c|c|c|}
	\hline
$n$     & $r=0.25$     & $r=0.30$     & $r=0.35$     \\
\hline \hline
$n=1$   & $L_0/a >10$  &$L_0/a >8.3$  &$L_0/a >7.1$  \\
$n=2$   & $L_0/a >18$  &$L_0/a >15$   &$L_0/a >12.8$ \\
$n=3$   & $L_0/a >26$  &$L_0/a >21.6$ &$L_0/a >18.5$ \\
	\hline
\end{tabular}
	\caption{The lower bound for $L_0/a$ to avoid over smearing.}\label{table:smearing-level}
	\end{center}
	\end{table}
%%%%%%%%%%%%%%%%%%%%

\begin{figure}[h]
   \begin{center}
      \unitlength=1mm
   \begin{picture}(105,75)
     \put(-2,72){\large $(\hat{R}+1/2)^2 \cdot \chi$}
     \put(49,-2){\large $\hat{R}+1/2$}
     \includegraphics[height=7cm]{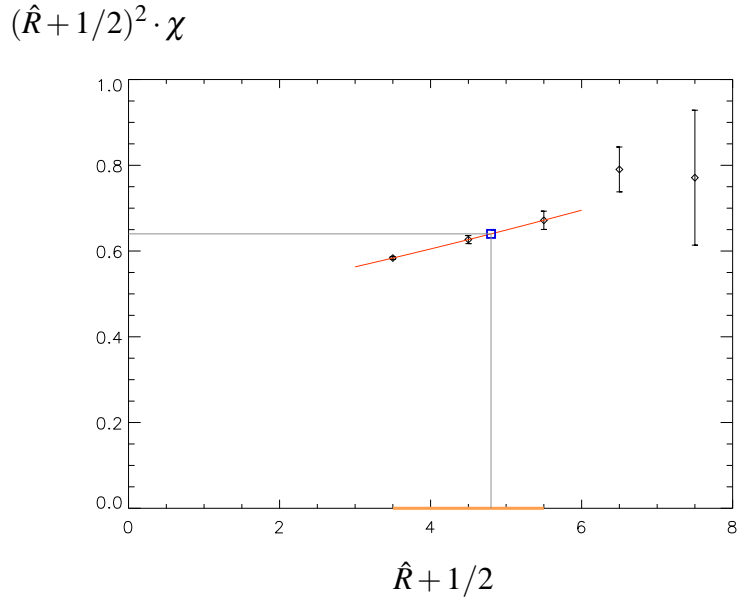}
     \end{picture}
   \end{center}
 \caption{The plot for the interpolation of $(\hat{R}+1/2)^2 \cdot \chi$ in the case of $\beta = 6.5700$ 
 and $L_0/a = 16$. The diamonds show the resultant values of $(\hat{R}+1/2)^2 \cdot \chi$ for integer $\hat{R}$ with statistical error. The red line represents the function $f(\hat{R}+1/2)$ fitted to the data in the range of $3 \le \hat{R} \le 5$. The blue box indicates the interpolated value corresponding to 
 $r = 0.3$.}
 \label{fig:Interpolation}
 \end{figure}

From here, we fix the renormalization condition $r=0.3$, then we have to calculate the 
value of $(\hat{R}+1/2)^2 \cdot \chi$ for noninteger $\hat{R}$.
We interpolate the value of $(\hat{R}+1/2)^2 \cdot \chi$ using a quadratic fit function:
\beq
f(\hat{R}+1/2)=c_0 +c_1 (\hat{R}+1/2) +c_2(\hat{R}+1/2)^2.
\eeq
We determine the fit ranges for each lattice size as in Table~\ref{table:fit-range}. 
We confirmed that the data can be well fitted by our fit function with these fit ranges 
for all parameter sets. An example is shown in Fig.~\ref{fig:Interpolation}.

%%%%%%%%%%%%%%%%%% Table %%%%%%%%%%%%%%%%%%%%%%
\begin{table}[h]
	
	\begin{center}
	\begin{tabular}{|c|c|c|c|}
	\hline
$L_0/a$ & $\hat{R}+1/2$ & $\hat{R}_{min}$ & $\hat{R}_{max}$     \\
\hline \hline
$12$   & $3.6$  &$2$  &$4$  \\
$14$   & $4.2$  &$2$  &$5$ \\
$16$   & $4.8$  &$3$  &$5$ \\
$18$   & $5.4$  &$4$  &$6$  \\
$20$   & $6.0$  &$4$  &$6$ \\
$24$   & $7.2$  &$5$  &$7$ \\
	\hline
\end{tabular}
	\caption{The fit range used to interpolate the value of $(\hat{R}+1/2)^2 \cdot \chi$. 
	The column ``$\hat{R}+1/2$'' is the value that corresponds to  $r=0.3$.}
	\label{table:fit-range}
	\end{center}
	\end{table}
%%%%%%%%%%%%%%%%%%%%

%%%%%%%%%%%%%%%%%%%%
%%%%%%%%%%%%%%%%%%%%

\begin{figure}[h]
\begin{center}
\unitlength=1mm
\begin{picture}(105,75)
\put(0,40){$g_w^2$}
\put(49,-2){$(a/L_0)^2$}

\includegraphics[height=7cm]{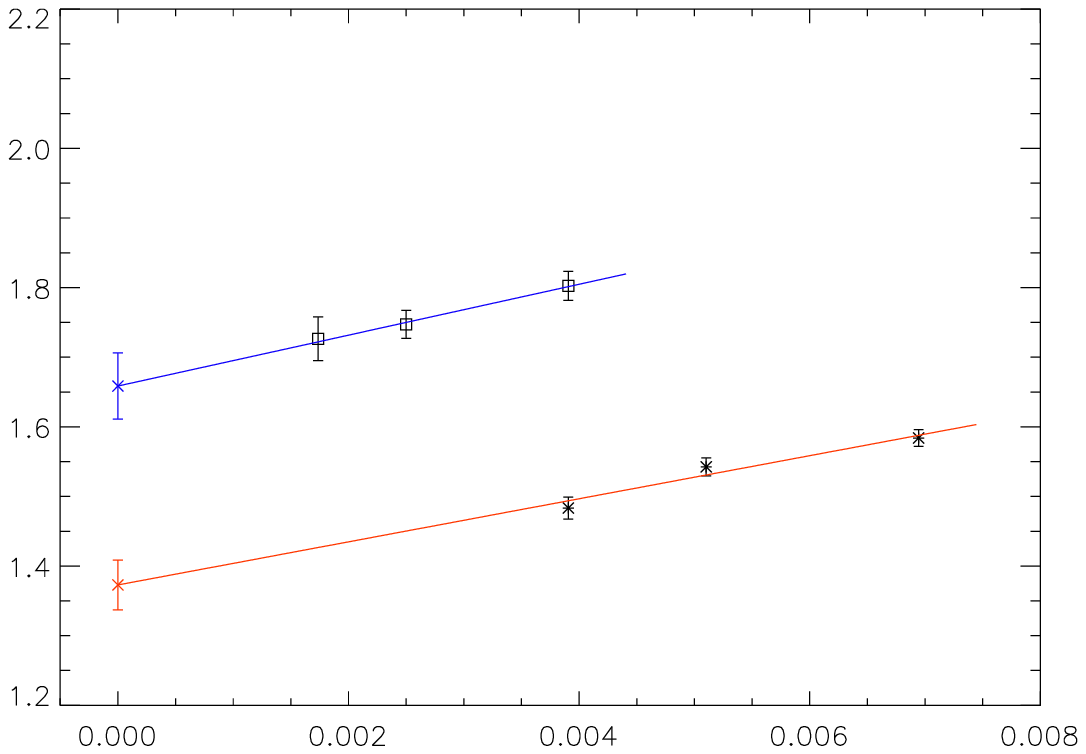}

\end{picture}
\end{center}
\caption{The continuum limit of $g_w^2$ in Set~$1$. The red and blue lines describe the continuum limit of the first and second columns of Set~$1$, respectively.}
\label{fig:Cont-Limit}
\end{figure}

Finally, we take the continuum limit of the running coupling constant of each column in Table~\ref{table:parameter-set}.
We show the example of the extrapolation of the coupling constant in Fig.~\ref{fig:Cont-Limit}.
The red line corresponds to the extrapolation of the first column of Set~$1$, 
and the blue line is that of the second column of Set~$1$.
The values at the continuum limit are the renormalized coupling constants at scales 
$1/\tilde{L}_0$ and $1/(2\tilde{L}_0)$, respectively.
In performing the continuum extrapolation, we have used the fit function 
linear in $(a/L_0)^2$:
\beq
c_0+c_1\left( \frac{a}{L_0} \right)^2.
\label{eq:linear-fit}
\eeq 
As explained in section $2$, our Wilson loop scheme 
does not have $O(a)$ systematic error.
However, in the present quenched QCD test, 
since we took the SF coupling, which has $O(a)$ error, 
as a constant input parameter,   
it is possible that an $O(a)$ error is introduced also in our data.
Fig.~\ref{fig:Cont-Limit} shows that the data are nicely fitted by the 
linear function, Eq.~(\ref{eq:linear-fit}). 
This would indicates that $O(a)$ systematic error and 
possible higher order discretization error are small as we 
expected from perturbative analysis \cite{Luscher:1996vw} 
in this weak coupling regime.

%%%%%%%%%%%%%%%%%%%%%%%
%%%%%%%%%%%%%%%%%%%%%%

%%%%%%%%%%%%%%%%%%%%%%%%%%%%%%%%%%
\section{Results and discussion}
\begin{figure}[h]
\begin{center}
\unitlength=1mm
\begin{picture}(105,75)
\put(0,40){\large{$g_w^2$}}
\put(78.2,3.7){\vector(0,1){4}}
\put(77.2,-0.7){$1$}
\put(67.5,3.7){\vector(0,1){4}}
\put(66.5,-0.7){$\frac{1}{2}$}
\put(56.9,3.7){\vector(0,1){4}}
\put(55.9,-0.7){$\frac{1}{4}$}
\put(46.7,3.7){\vector(0,1){4}}
\put(45.7,-0.7){$\frac{1}{8}$}
\put(36.2,3.7){\vector(0,1){4}}
\put(34.4,-0.7){$\frac{1}{16}$}
\put(25.7,3.7){\vector(0,1){4}}
\put(24.0,-0.7){$\frac{1}{32}$}
\put(88,-1){\large $\tilde{L}_0/L_0$}
\includegraphics[height=7cm]{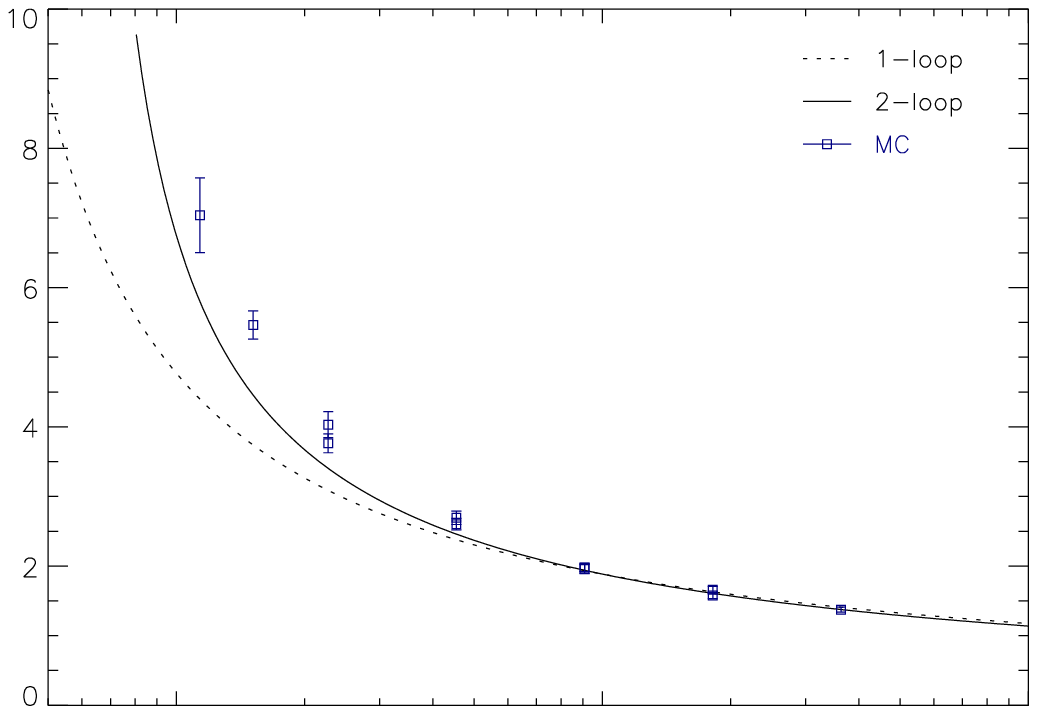}
\end{picture}
\end{center}
\caption{The running coupling in the 
Wilson loop scheme.  
The horizontal axis is the ratio of the energy scale $\tilde{L}_0/L_0$, 
and the vertical axis is the 
renormalized coupling in the Wilson loop scheme.
The square symbols indicate our lattice data of the running coupling constants. 
The right-most symbol corresponds to 
the renormalized coupling given by the first column of Set~$1$ at scale $1/{\tilde{L}}_0$,  
the adjacent two overlapping symbols give the couplings from the second column of Set~$1$ 
and the first column of Set~$2$ at scale $1/(2{\tilde{L}}_0)$, and similarly for the remaining 
data for lower energy scales.
The perturbative running couplings at one-loop (dotted line) and two-loop (solid line) 
are also shown for comparison.}
 \label{fig:run-coupling}
 \end{figure}

We plot the numerical results for the energy scale dependence of the renormalized coupling 
in the Wilson loop scheme in Fig.~\ref{fig:run-coupling}. 
The perturbative renormalization group evolution at one-loop and two-loop are also 
displayed in the figure for comparison. 
Our MC data are consistent with the perturbative results in high energy region, while there is 
difference in low energy region because of the higher loop or the nonperturbative effect.

Recall that the second column ($s=2$) of Set $1$ (or $2$, $3$, $4$) and 
the first column ($s=1$) of Set $2$ (or $3$, $4$, $5$) in 
Table~\ref{table:parameter-set} 
give the same value of the renormalized coupling in the SF scheme within the statistical error. 
From Fig.~\ref{fig:run-coupling}, we see that our results in the Wilson loop scheme also show that 
the step scaling procedure works well; steps are nicely connected as in the case of the SF couplings.
It suggests that the systematic errors, such as the interpolation in $\hat{R}+1/2$, $O(a)$ error 
in SF coupling and the extrapolation to the continuum limit are under control.
To make our study self-contained, we should extend this scheme by taking 
$g_w^2$,  which has no $O(a)$ systematic error, as an input physical 
parameter. Such study is underway, and will be presented 
in our future publication~\cite{k-paper}.

Finally, we emphasize that there is an optimal choice of smearing level and 
$r$ with which both the statistical and discretization errors are under control. 
The number of gauge configurations we used for the present study was only $100$ 
for each $(\beta, L_0/a)$. 
From this small number of gauge configurations, we obtained data with a few 
($\sim10$) percent 
statistical error in high (low) energy region.
Thus, we could expect that this scheme is promising for the study of other theories.

%%%%%%%%%%%%%%%%%%%%%%%
%%%%%%%%%%%%%%%%%%%%%%

%%%%%%%%%%%%%%%%%%%%%%%%%%%%%%%%%%
\section{Summary}

We proposed a new scheme for the determination of the running coupling 
on the lattice. Our method is based on the measurement of the finite volume 
dependence of the Wilson loop. Unlike the SF scheme, our method does 
not have any $O(a)$ discretization error, therefore the systematic error arising 
from the extrapolation to the continuum limit is expected to be quite small. 
We showed results of preliminary numerical study for the quenched QCD 
as a feasibility test of our scheme, and confirmed that the method 
actually reproduced the step scaling of the coupling which is consistent 
with the two-loop 
running coupling  at high energy. We also showed that the coupling 
calculated by this newly proposed scheme deviates from that with two-loop 
approximation below a certain energy scale. 
This deviation arises from the effects that are not captured by 
the two-loop approximation. We have confirmed that our scheme  
works well for the calculation of the running coupling with relatively 
small number of gauge configurations, via demonstrating that the statistical error 
is under control by properly choosing the smearing level and $r$. 
We expect that this new method is  quite useful to calculate the running coupling of the $SU(N)$ 
gauge theory with a large number of dynamical fermions, which will be 
studied in our future work.

The properties of a vectorial gauge theory as a function of the 
number of fermions is quite interesting. Through the analyses based 
on the Schwinger-Dyson and Bethe-Salpeter equations, it is shown 
that there is chiral symmetry restoration at certain critical number of fermions 
\cite{chipt2, chipt3}, and that the behavior of the meson spectrum, the $S$ parameter, 
\textit{etc.}, are quite different from those in QCD with three flavors. 
\cite{Harada:2005ru, Kurachi:2006mu, Harada:2003dc, Kurachi:2007at}.
In these analyses, two-loop running coupling was used as an approximation 
to the fully non-perturbative vertices. Therefore, the existence of the 
IRFP is an assumption and an input in these analyses. The confirmation 
of the existence of the IRFP by lattice simulations justifies this assumption,  
and motivates a fully 
non-perturbative study of near conformal gauge theories on the lattice.

\section*{Acknowledgment}
This work is supported in part by the Grant-in-Aid of the Ministry of Education
(Nos. 19540286, 19740160 and 20039005).
E.~B. is supported by the FWF Doktoratskolleg Hadrons in Vacuum,
Nuclei and Stars (DK W1203-N08). 
A.~F. acknowledges the support of JSPS, Grant N.19GS0210.
C.-J~D.~L. is supported by the National Science Council of Taiwan via
grant number 96-2112-M-009-020-MY3. 
H.~O. thanks N. Yamada for his support, especially by Grand-in-Aid for 
Scientific Research No. 20025010 from the Ministry of Education, Culture, Sports, 
Science and Technology of Japan. 
T.Y. is the Yukawa Fellow supported by Yukawa Memorial Foundation. 
Numerical simulation was carried out on the vector supercomputer NEC SX-8 in 
YITP, Kyoto University.

%%%%%%%%%%%%%%%%%%%%%%%%%%%%%%%%%%

\end{document}